# Towards large-area and defects-free growth of phosphorene on Nickel.

B. D. Tchoffo,[1,+] I. Benabdallah,[1,*,+] A. Aberda,[1] P. Neugebauer,[2] A. Belhboub,[3] A. El Fatimy[1*]

[1] Institute of Applied Physics, Mohammed VI Polytechnic University, Lot 660, Hay Moulay Rachid, Ben Guerir, 43150, Morocco.

[2] Central European Institute of Technology, CEITEC BUT, Purkyňova 656/123, 61200 Brno, Czech Republic.

[3] École Centrale Casablanca, Ville verte, Bouskoura, Casablanca, Morocco.

[+]These authors contributed equally

[*]Corresponding author: ismail.benabdallah@um6p.ma, Abdelouahed.ELFATIMY@um6p.ma

Low-dimensional materials synthesis based on phosphorus atoms is under intense study, and it is still one of the big challenges. Phosphorene, a monolayer of black phosphorus, is one of the most promising candidates for transistor and photonics devices at atomistic thickness. However, the lack of large-scale and defects-free growth significantly obstructs its device development. Here, we demonstrate the large-scale and defect-free phosphorene synthesis on Nickel (Ni) substrate. In addition, the effect of substrate orientation on the controllable synthesis of possible allotropes has also been described. We have shown that blue phosphorene can be grown on Ni (111) and Ni (100). While γ-Phosphorene, named Navy Phosphorene hereafter, can be grown on Ni (110). Furthermore, we found that the synthesis goes through phosphorus pentamers (P5) to phosphorene; P5 is a vital precursor for phosphorene synthesis. Moreover, we confirm the high accuracy of the P-Ni, and P-P potentials and show that the molecular dynamics (M.D.) approach is a powerful tool to simulate the 2D materials synthesis in the vapor phase. This work provides a solid reference to understand and control the synthesis of large-area single-crystalline monolayer phosphorene.

The discovery of two-dimensional (2D) materials based on phosphorus atoms (phosphorene) initiated a plethora of research and development activities on this novel material due to its enthralling semiconducting properties.[10-12]. Many allotropic forms of phosphorene have been investigated using first principle studies and were predicted to be stable structural phases. These allotropes include black phosphorene (α-P),[1] blue phosphorene (β-P),[2] green phosphorene,[3] δ-phosphorene,[1] navy Phosphorene (γ-P),[12,] and others.[7,22] Among these allotropes, the blue phosphorene has a similar honeycomb lattice from the top view as graphene. In addition, blue phosphorene demonstrates isotropic optical and electronic properties with an indirect bandgap of 2.0 eV and an ultrahigh carrier (holes) mobility up to $10^3$ $cm^2V^{-1}s^{-1}$.[12,23] On the other hand, navy phosphorene (γ-P) exhibits an anisotropic structure, with interesting anisotropic optical properties, an indirect band gap of 1.03 eV, and also the highest carrier (electrons) mobility up to $10^5$ $cm^2V^{-1}s^{-1}$ as compared to other phosphorene allotropes.[12]

In preliminary results, the above allotropes have been successfully exploited in several applications, such as electronics,[8] optoelectronics,[14,15] spintronics,[13,15] sensors,[16] storage devices,[17] and terahertz applications,[18] exhibiting the high potential of these 2D allotropes of phosphorene. However, the lack of large-scale and defect-free synthesis is the major issue hindering the development and application of allotropic phosphorene in modern devices such as Terahertz detectors[19-21]. So far, the primary prototype devices have only been demonstrated based on exfoliated phosphorene from bulk black phosphorus as graphite to graphene. The controllable large-scale defect-free growth of monolayer phosphorene is essential for the properties mentioned earlier. Although, tiny-size blue phosphorene has been synthesized using molecular beam epitaxy on Au (111), Ag (111), and other metal oxide substrates.[4-6] However, a four-inch wafer-scale uniform monolayer is essential for developing phosphorene-based devices. Nowadays, Chemical

Vapor Deposition (CVD) synthesis is the most promising method for synthesizing large-area, high-quality monolayer 2D materials such as graphene, TMDC, etc.[8,9] Owing to its ability to explore the vast parameters of the substrate, precursors, temperature, pressure, etc., may enable the bottom-up routes for large-scale phosphorene synthesis. To date, the investigations of CVD synthesis of large-area and uniform phosphorene have yet to be reported. In this work, we propose to use the simulation methods based on molecular dynamics (M.D.) to study the 2D materials synthesis in the vapor phase. M.D. simulations were carried out using the LAMMPS package, which implements the Verlet algorithm for integrating Newton's equations in time.[24,25] A time step of 0.25 fs was used to perform the integrations as it is small enough to obtain accurate results for temperatures below 2500 K.[26] The simulation cell (surface area of about 1400Å² in each case) is comprised of a Ni substrate oriented along the (100), (110), and (111) lattice planes.

Here, we demonstrate the synthesis of blue and navy phosphorene on Ni. We also show that Ni (111) and Ni (100) orientation exhibit the lowest defects due to symmetry and lattice mismatch for the blue phosphorene. Furthermore, we found that navy phosphorene synthesis is possible on Ni (110) due to the corrugated surface of the substrate. It was also found that the phosphorene synthesis process can be divided into three major phases and the phosphorus pentamer ($P_5$) is the vital precursor to phosphorene growth on Ni. This demonstrates that a large and almost perfect blue and navy phosphorene layer can be grown on Ni. Several parameters were carefully studied to understand better the phosphorus deposition process: substrate orientation, temperature, and cooling rate. To approximate the vapor deposition process, the temperature was raised from 300 to 1700 K (~ Nickel melting temperature) during a period of 0.25 ps; then it was held constant in the maximum temperature plateau during 150 ps before being cooled to 300 K for 100 ps, as shown in the temperature profile in Fig. 1 (g).

Fig.1 shows the simulated surface formed by the phosphorus atoms deposited on three different orientations of the nickel substrate, i.e. (100), (110), and (111). Fig. 1 (a), (c), and (e) illustrates the evolution of phosphorene on the surface of the substrate at 1500 K (Ni (100) and Ni (110)) and 1700 K (Ni (111)). Fig. 1 (b), (d), and (f) visibly demonstrate that more defects are present before cooling down to 300 K for each orientation of the substrate. From this, it could be understood that the cooling rate plays an essential role in allowing the phosphorus atoms to be transported from the aggregated areas to the vacancy sites. Yu et al. also reported a similar phenomenon indicating that segregation highly depends on the cooling rate.[27] The optimal growth temperatures were found to be 1500 K for (100) and (110); 1700 K for (111) orientations, as lower dislocations and defects could be seen after cooling as compared to other temperatures (other temperature profiles are shown in figure S1). By comparing the structures obtained for each substrate orientation at the end of the cooling phase, it can be seen that the (111) orientation has a better structure than the other orientations. This suggests that Ni (111) is the most stable orientation for the growth of a homogeneous phosphorene layer.

Fig. 2 shows the complete simulation profile for different cooling rates ranging from 2.8 K/ps to 0.116 K/ps with different time steps. For the cooling rate of 0.116 K/ps, an almost defect-free phosphorene layer was formed on each considered orientation of the Nickel substrate. On the other hand, single vacancies (circled in blue), double vacancies (circled in green), and Stone Wales defects (circled in red) are present for fast cooling rates. It is because fast cooling causes a quenching effect, preventing atoms from being transported to the vacancies, contrary to slow cooling. Apart from the defect-free and uniform layer obtained with the slow cooling, we show the importance of substrate orientation since substrate orientation can be used to select the type of phosphorene allotropes formed. For example, navy phosphorene (γ-phosphorene) was obtained on Ni (110), while blue phosphorene was formed on Ni (100) and Ni (111).

Fig. 3 shows the detailed configuration of the system for Ni orientations at each time step for fast and slow cooling of 2.8 K/ps and 0.116 K/ps, respectively. Fig. 3 (a) depicts the $P_n$ evolution for Ni (100), which shows a non-decreasing concentration of $P_5$, which explains the higher defect concentrations found above for phosphorene on Ni (100). For the other orientations (110) and (111), as shown in Fig. 3 (b), (c) a noticeable decrease is observed for $P_5$; while $P_7$ clusters are increasing, this behavior can be understood by the presence of Stones-Wales defects type (5-7-7-5). These defects are produced by rotating one phosphorus atom pair, resulting in four $P_6$ hexagonal rings (6-6-6-6) converting to two $P_5$ and $P_7$ rings.[28] Additionally, there are also some $P_8$ rings related to double vacancy defects of type (5-8-5)[30]. Still, Stone-Wales defects are the most prominent due to their low energy formation of ~1.60 eV compared to single or double vacancies of ~ 2.85 eV.[31] These defects are observed in other 2D materials synthesis due to fast cooling [29]. Furthermore, we show that the phosphorene synthesis process can be divided into three major phases. The first phase corresponds to the atomic deposition phase. In the second phase, nucleation centers emerge with atoms forming *n*-membered rings ($P_n$). Several $P_n$ clusters, such as $P_4$, $P_8$, and $P_7$, are unstable and disappear as soon as they form. Therefore, additional $P_n$ are formed simultaneously, such as $P_3$, $P_5$, and $P_6$. The last phase corresponds to the cooling process. All the $P_n$ forms disappear or decrease except for $P_5$, $P_6$, and $P_7$, which converge to a constant value over time (Descriptive videos can be found in the supplementary information showing these processes).

Our Physical picture of the growth mechanism is as follows: initially, $P_5$ clusters are formed due to their low formation energy; upon reaching a high surface coverage on the substrate, they convert to $P_6$; hence, forming the hexagonal puckered honeycomb of blue and navy phosphorene. Thus, suggesting that $P_5$ is the natural precursor to synthesize phosphorene allotropes on Ni.

Fig. 4 shows the radial distribution function (R.D.F.) and the Angular Distribution Function (A.D.F.) of the deposited layer on the Ni substrate for a cooling rate of 0.116 K/ps. Figure 4 (a) shows the β-P RDF with four significant peaks, quite similar behavior to Ni (111) and Ni (100). On the other hand, the RDF of γ-P shows five significant peaks, with an additional peak at 4.84Å.

| References | Bond length (Å) | Lattice parameter (Å) | Angle (°) |
|---|---|---|---|
| **DFT calculations at 0 K (β-P)** [2, 33-36] | 2.26 – 2.27 | 3.28 – 3.33 | 92.9 – 93.07 |
| **DFT calculations at 0 K (γ-P)** [34] | 2.25 – 2.31 | $a = 3.42; b = 5.27$ | 93.25 – 100.21 |
| **M.D. simulation at 0 K (β-P)** [37] | 2.24 | 3.45 | - |
| **M.D. simulation at 0 K (γ-P)** [37] | 2.215; 2.312 | $a = 3.407; b = 5.793$ | - |
| **The present study at 300 K (β-P)** | 2.25 | 3.6 | 102.5 – 103.5 |
| **The present study at 300 K (γ-P)** | 2.23 | $a = 3.534; b = 5.48$ | 103.5 |

**Table 1:** Optimized structure parameters of the blue phosphorene sheets at zero strain at 0 K for previous studies and 300 K for the present study

The first peak of the RDF of each allotrope is located approximately at the same positions corresponding to the P-P bond's length; this value is consistent with other theoretical works.[37] The second peak is the lattice parameter of β–P and γ–P, which are slightly different, as shown in Table. 1. For the γ–P, the last peak at 5.48 Å is related to the lattice parameter *b*. The other unmentioned peaks arise from the third and fourth nearest neighbors. Figure 4 (b) shows the mean angle between the P-P-P bond; the obtained values are close to each other for the different orientations. The lattice parameters and angles are slightly different compared to the DFT calculation. The influence of temperature and the substrate's effect may explain these differences.

In summary, we report the synthesis strategy to grow high-quality, large-scale navy and blue phosphorene on a Ni (110), (100), and (111) substrate. In addition, we show that substrate orientation can be used to control and select the synthesis of allotropes. Furthermore, we found that the synthesis goes through phosphorus pentamers ($P_5$) to phosphorene; $P_5$ is a vital precursor for phosphorene synthesis. This preliminary work provides a solid reference to understand and control the synthesis of two-dimensional materials based on phosphorus atoms.

**Methods:**

The simulation box used for M.D. simulations contains the Ni substrate on which the phosphorus atoms are deposited. Each substrate contains six layers stacked along the z-axis with a surface area of about 1400 Å² each. The two bottom layers of the substrate were fixed throughout the simulation. A vacuum of 40 Å was kept on top of the first layer of the substrate to allow the phosphorus atoms to deposit freely. To reduce the boundary effects, periodic boundary conditions were applied in the x and y directions, while the z-axis had a finite dimension to define a fixed box size. Additionally, a wall reflect was applied at the upper side of the simulation box to prevent atoms from leaving the box.

There are two types of atoms in the simulation box, Phosphorus (P) and Nickel (Ni), with three kinds of interactions: Ni-Ni, Ni-P, and P-P. These interactions are described by two different interatomic potentials, one of which is the well-known Lennard-Jones (L.J.) potential given by eq. (1).

$$U(r) = 4\varepsilon \left[ \left(\frac{\sigma}{r}\right)^{12} - \left(\frac{\sigma}{r}\right)^{6} \right] \quad \text{(eq. 1)}$$

Where ε and σ are the L.J. parameters and are characteristic of each atom; ε represents the strength of the interaction, and σ is the distance parameter.

Herein the L.J. potential describes the Ni-Ni and Ni-P interactions. Knowing that Ni-P is an interaction between two different atoms, the combining rule formula given by eq. (2) was used to estimate the L.J. parameters in this case.

$$\varepsilon = (\varepsilon_{Ni}.\varepsilon_P)^{1/2} \quad ; \quad \sigma = \left(\frac{\sigma_{Ni}+\sigma_P}{2}\right) \quad (eq.2)$$

The L.J. parameters are given for each atom in Table (2).

**Table. 2**: Lennard-Jones parameters used for the simulations

|    | $\varepsilon\ (kcal/mol)$ | $\sigma(\text{Å})$ |
|----|---------------------------|--------------------|
| Ni | 8.599276[38]              | 2.2808[38]         |
| P  | 0.500000[39]              | 3.3300[39]         |

Knowing that many chemical processes occur during the CVD synthesis, such as bond breaking and bond creation, it is not appropriate to use a non-reactive force field to describe the P-P interactions as the bonds could be broken one timestep and created the next. For these reasons, it is more appropriate to use a force field that can describe all the chemical mechanisms during the simulation, hence, the use of the ReaxFF potential (Reactive Force Field).[40] It was also found that ReaxFF can provide energies, transition states, reaction paths, and reactive events that are consistent with quantum mechanical calculation.[41] For this study, a recent version developed by Xiao et *al.* was used as it gives a good description of the physical and mechanical properties of phosphorus and its allotropes.[40]

Before deposition, the box containing the substrate was first minimized. For all the simulations, 280 atoms were deposited. This value was chosen as it would be sufficient to form a whole layer of phosphorene on the chosen substrate surface. The substrate was first rapidly heated from 300K to a higher temperature (ranging from 400 K to 1700 K) before the phosphorus atoms were deposited. Following the heating treatment, the system was cooled from the growth temperature back to room temperature with different cooling rates to observe the cooling effect on the synthesis process. The setup to make the simulation approximate the CVD synthesis method would be depositing $P_4$ phosphorus gas molecules corresponding to red phosphorus. However, we only use single phosphorus atoms to be deposited on the Nickel surface as the $P_4$ clusters would dissociate as the temperature increases. Here, major synthesis parameters were investigated to study the mechanism of phosphorus deposition, such as substrate orientations, temperature, deposition time, and cooling rate.

**Acknowledgments:** OCP Foundation has supported this work with the project grant AS70, "Towards phosphorene-based materials and devices, and with the support of the Chair "Multiphysics and HPC" led by Mohammed VI Polytechnic University. The ERC supported this work under the European Union's Horizon 2020 research and innovation program (GA No. 714850). We acknowledge the High-Performance Computing (HPC) Facility of Mohammed VI Polytechnic University – Toubkal.

**References:**

1. Zhu, Z., & Tománek, D. Semiconducting layered blue phosphorus: a computational study. *Physical review letters* **112(17)**, 176802 (2014).

2. Han, W. H., Kim, S., Lee, I. H., & Chang, K. J. Prediction of green phosphorus with tunable direct band gap and high mobility. *The journal of physical chemistry letters* **8(18)**, 4627-4632 (2017).

3. Zhang, W. et al. Epitaxial synthesis of blue phosphorene. *Small* **14.51**, 1804066 (2018).

4. Zhuang, J. et al. Band gap modulated by electronic superlattice in blue phosphorene. *ACS nano* **12.5**, 5059-5065 (2018).

5. Yang, S. et al. Regular Arrangement of Two-Dimensional Clusters of Blue Phosphorene on Ag (111). *Chinese Physics Letters* **37.9**, 096803 (2020).

6. Zhao, T. et al. A new phase of phosphorus: the missed tricycle type red phosphorene. *Journal of Physics: Condensed Matter* **27.26**, 265301 (2015).

7. Chaudhary, V., Neugebauer, P., Mounkachi, O., Lahbabi, S., & EL FATIMY, A. Phosphorene-an emerging two-dimensional material: recent advances in synthesis, functionalization, and applications. *2D Materials*, (2022).

8. El Hammoumi, M., Chaudhary, V., Neugebauer, P., & El Fatimy, A. Chemical vapor deposition: a potential tool for wafer scale growth of two-dimensional layered materials. Journal of Physics D: Applied Physics, 55(47), 473001 (2022).

9. Liu, H. et al. Phosphorene: an unexplored 2D semiconductor with a high hole mobility. *ACS Nano* **8**, 4033–4041 (2014).

10. Xia, F., Wang, H. & Jia, Y. Rediscovering black phosphorus as an anisotropic layered material for optoelectronics and electronics. *Nat. Commun* **5**, 4458 (2014).

11. Castellanos-Gomez, A. Black phosphorus: narrow gap, wide applications. *The journal of physical chemistry letters* **6(21)**, 4280-4291 (2015).

12. Xiao, J. et al. Theoretical predictions on the electronic structure and charge carrier mobility in 2D phosphorus sheets. *Scientific reports* **5(1)**, 1-10 (2015).

13. Guo, Z. et al. Metal-ion-modified black phosphorus with enhanced stability and transistor performance. *Advanced Materials* **29(42)**, 1703811 (2017).

14. Buscema, M. et al. Fast and broadband photoresponse of few-layer black phosphorus field-effect transistors. *Nano letters* **14(6)**, 3347-3352 (2014).

15. Kamalakar, M. V., Madhushankar, B. N., Dankert, A., & Dash, S. P. Engineering Schottky barrier in black phosphorus field effect devices for spintronic applications. *arXiv preprint arXiv:1406.4476* (2014).

16. Zhu, Y. et al. From phosphorus to phosphorene: Applications in disease theranostics. *Coordination Chemistry Reviews* **446**, 214110 (2021).

17. Pang, J. et al. Applications of phosphorene and black phosphorus in energy conversion and storage devices. *Advanced Energy Materials* **8(8)**, 1702093 (2018).

**References:**

1. Zhu, Z., & Tománek, D. Semiconducting layered blue phosphorus: a computational study. *Physical review letters* **112(17)**, 176802 (2014).

2. Han, W. H., Kim, S., Lee, I. H., & Chang, K. J. Prediction of green phosphorus with tunable direct band gap and high mobility. *The journal of physical chemistry letters* **8(18)**, 4627-4632 (2017).

3. Zhang, W. et al. Epitaxial synthesis of blue phosphorene. *Small* **14.51**, 1804066 (2018).

4. Zhuang, J. et al. Band gap modulated by electronic superlattice in blue phosphorene. *ACS nano* **12.5**, 5059-5065 (2018).

5. Yang, S. et al. Regular Arrangement of Two-Dimensional Clusters of Blue Phosphorene on Ag (111). *Chinese Physics Letters* **37.9**, 096803 (2020).

6. Zhao, T. et al. A new phase of phosphorus: the missed tricycle type red phosphorene. *Journal of Physics: Condensed Matter* **27.26**, 265301 (2015).

7. Chaudhary, V., Neugebauer, P., Mounkachi, O., Lahbabi, S., & EL FATIMY, A. Phosphorene-an emerging two-dimensional material: recent advances in synthesis, functionalization, and applications. *2D Materials*, (2022).

8. El Hammoumi, M., Chaudhary, V., Neugebauer, P., & El Fatimy, A. Chemical vapor deposition: a potential tool for wafer scale growth of two-dimensional layered materials. Journal of Physics D: Applied Physics, 55(47), 473001 (2022).

9. Liu, H. et al. Phosphorene: an unexplored 2D semiconductor with a high hole mobility. *ACS Nano* **8**, 4033–4041 (2014).

10. Xia, F., Wang, H. & Jia, Y. Rediscovering black phosphorus as an anisotropic layered material for optoelectronics and electronics. *Nat. Commun* **5**, 4458 (2014).

11. Castellanos-Gomez, A. Black phosphorus: narrow gap, wide applications. *The journal of physical chemistry letters* **6(21)**, 4280-4291 (2015).

12. Xiao, J. et al. Theoretical predictions on the electronic structure and charge carrier mobility in 2D phosphorus sheets. *Scientific reports* **5(1)**, 1-10 (2015).

13. Guo, Z. et al. Metal-ion-modified black phosphorus with enhanced stability and transistor performance. *Advanced Materials* **29(42)**, 1703811 (2017).

14. Buscema, M. et al. Fast and broadband photoresponse of few-layer black phosphorus field-effect transistors. *Nano letters* **14(6)**, 3347-3352 (2014).

15. Kamalakar, M. V., Madhushankar, B. N., Dankert, A., & Dash, S. P. Engineering Schottky barrier in black phosphorus field effect devices for spintronic applications. *arXiv preprint arXiv:1406.4476* (2014).

16. Zhu, Y. et al. From phosphorus to phosphorene: Applications in disease theranostics. *Coordination Chemistry Reviews* **446**, 214110 (2021).

17. Pang, J. et al. Applications of phosphorene and black phosphorus in energy conversion and storage devices. *Advanced Energy Materials* **8(8)**, 1702093 (2018).

18. Fang, R. et al. Strain-Engineered ultrahigh mobility in Phosphorene for Terahertz Transistors. *Advanced Electronic Materials* **5(3)**, 1800797 (2019).

19. Tsuda, Y. et al. application of plasmon-resonant microchip emitters to broadband terahertz spectroscopic measurement. JOSA B 26.9, A52-A57 (2009).

20. El Fatimy, A. et al. "Effect of defect-induced cooling on graphene hot-electron bolometers." Carbon 154, 497-502 (2019).

21. El Fatimy, A. et al. Epitaxial graphene quantum dots for high-performance terahertz bolometers. Nature nanotechnology 11.4, 335-338 (2016).

22. Sun, H., Liu, G., Li, Q., & Wan, X. G. First-principles study of thermal expansion and thermomechanics of single-layer black and blue phosphorus. *Physics Letters A* **380(24)**, 2098-2104 (2016).

23. Xiao, J. et al. Electronic structures and carrier mobilities of blue phosphorus nanoribbons and nanotubes: a first-principles study. *The Journal of Physical Chemistry C* **120(8)**, 4638-4646 (2016).

24. Plimpton, S. Fast parallel algorithms for short-range molecular dynamics. *Journal of computational physics* **117(1)**, 1-19 (1995).

25. Verlet, L. Computer" experiments" on classical fluids. I. Thermodynamical properties of Lennard-Jones molecules. *Physical review* **159(1)**, 98 (1967).

26. Weismiller, M. R., Van Duin, A. C., Lee, J., & Yetter, R. A. ReaxFF reactive force field development and applications for molecular dynamics simulations of ammonia borane dehydrogenation and combustion. *The Journal of Physical Chemistry A* **114(17)**, 5485-5492 (2010).

27. Yu, Q. et al. graphene segregated on Ni surfaces and transferred to insulators. *Applied physics letters* **93(11)**, 113103 (2008).

28. Tian, W., Li, W., Yu, W., & Liu, X. A review on lattice defects in graphene: types, generation, effects and regulation. *Micromachines* **8(5)**, 163 (2017).

29. Wang, C., You, Y., & Choi, J. H. First-principles study of defects in blue phosphorene. *Materials Research Express* **7(1)**, 015005 (2019).

30. Sun, M., Chou, J. P., Hu, A., & Schwingenschlogl, U. Point defects in blue phosphorene. *Chemistry of Materials* **31(19)**, 8129-8135 (2019).

31. Wu, Z., & Ni, Z. Spectroscopic investigation of defects in two-dimensional materials. *Nanophotonics* **6(6)**, 1219-1237 (2017).

32. Qiu, L., Dong, J., & Ding, F. Selective growth of two-dimensional phosphorene on catalyst surface. *Nanoscale* **10(5)**, 2255-2259 (2018).

33. Swaroop, R., Ahluwalia, P. K., Tankeshwar, K., & Kumar, A. Ultra-narrow blue phosphorene nanoribbons for tunable optoelectronics. *R.S.C. advances* **7(5)**, 2992-3002 (2017).

34. Nahas, S., Bajaj, A., & Bhowmick, S. Polymorphs of two-dimensional phosphorus and arsenic: insight from an evolutionary search. *Physical Chemistry Chemical Physics* **19(18)**, 11282-11288 (2017).

35. Kaewmaraya, T. et al. Novel green phosphorene as a superior chemical gas sensing material. *Journal of Hazardous Materials* **401**, 123340 (2021).

36. Mogulkoc, Y., Modarresi, M., Mogulkoc, A., & Ciftci, Y. O. Electronic and optical properties of bilayer blue phosphorus. *Computational Materials Science* **124**, 23-29 (2016).

37. Le, M. Q. Reactive molecular dynamics simulations of the mechanical properties of various phosphorene allotropes. *Nanotechnology* **29(19)**, 195701 (2018).

38. Mauchamp, N. A., Ikuse, K., Isobe, M., & Hamaguchi, S. Self-sputtering of the Lennard–Jones crystal. *Physics of Plasmas* **29(2)**, 023507 (2022).

39. Sresht, V., Padua, A. A., & Blankschtein, D. Liquid-phase exfoliation of phosphorene: design rules from molecular dynamics simulations. *ACS nano* **9(8)**, 8255-8268 (2015).

40. Xiao, H. et al. Development of a transferable reactive force field of P/H systems: application to the chemical and mechanical properties of phosphorene. *The Journal of Physical Chemistry A* **121.32**, 6135-6149 (2017).

41. Mueller, J. E., Van Duin, A. C., & Goddard III, W. A. Development and validation of ReaxFF reactive force field for hydrocarbon chemistry catalyzed by Nickel. *The Journal of Physical Chemistry C* **114(11)**, 4939-4949 (2010).

**Fig 1.** (a) Top view of the final structure at 1500 K before and (b) after cooling on Ni (100). (c) Top view of the final structure at 1500 K before and (d) after cooling on Ni (110). (e) Top view of the final structure at 1700 K before and (f) after cooling on Ni (111); (g) Temperature profile used for the heating and cooling process.

**Fig 2.** Top views of the obtained layer at a cooling of 2.8 K/ps, 1.4 K/ps, 0.93 K/ps, 0.175 K/ps, and 0.116 K/ps for all the orientations. Circled areas represent ⭕ Stone Wales, ⭕ Single vacancy, and ⭕ Double vacancy defects.

**Fig 3.** The number of n-membered rings (Pn) in final configurations as a function of simulation time, for (a) 2.8 K/ps and (b) 0.116 K/ps for Ni (100); (c) 2.8 K/ps and (d) 0.116 K/ps for Ni (110); (e) 2.8 K/ps and (f) 0.116 K/ps for Ni (111). (The inset of Fig. 3 (b), (d), and (f) shows the full simulation corresponding to the cooling rate of 0.116 K/ps, while Fig. 3 (b), (d), and (f) is a zoom-in on the area of interest)

**Fig 4.** (a) Radial distribution function (RDF) and (b) angular distribution function (A.D.F.) of the formed layer for a cooling rate of 0.116 K/ps.

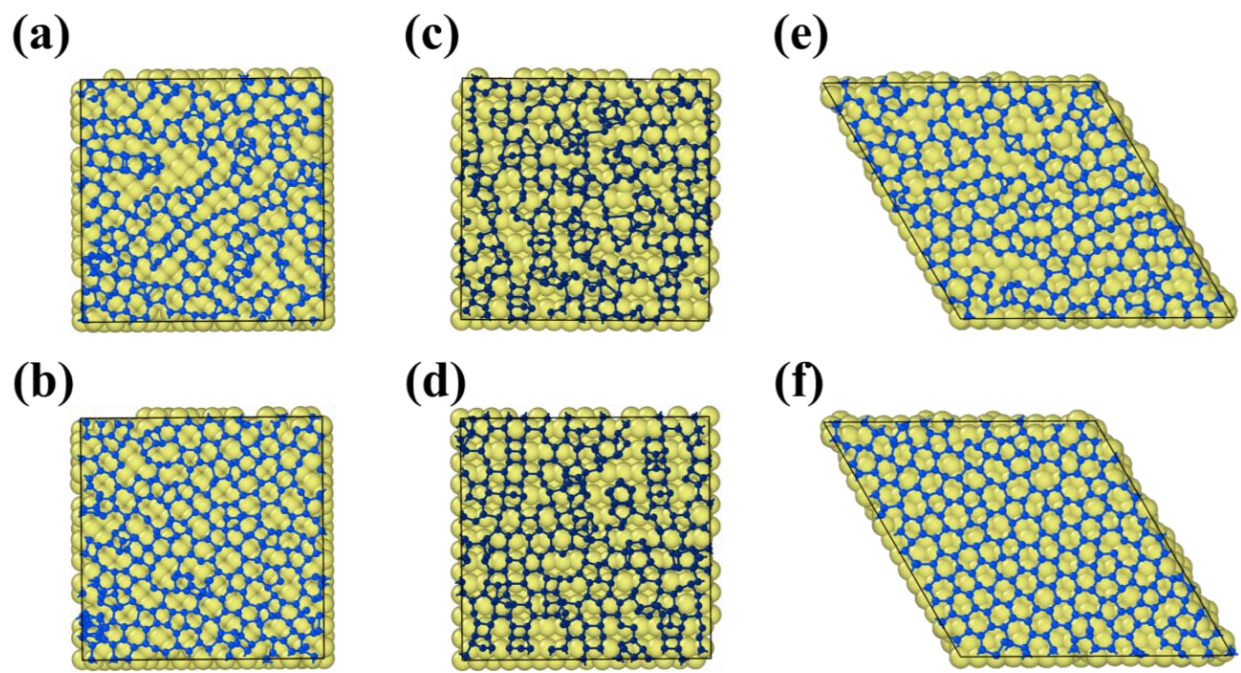
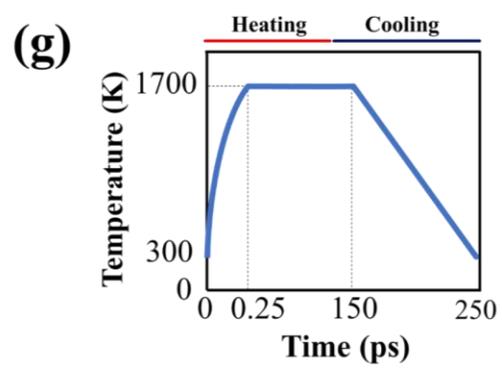

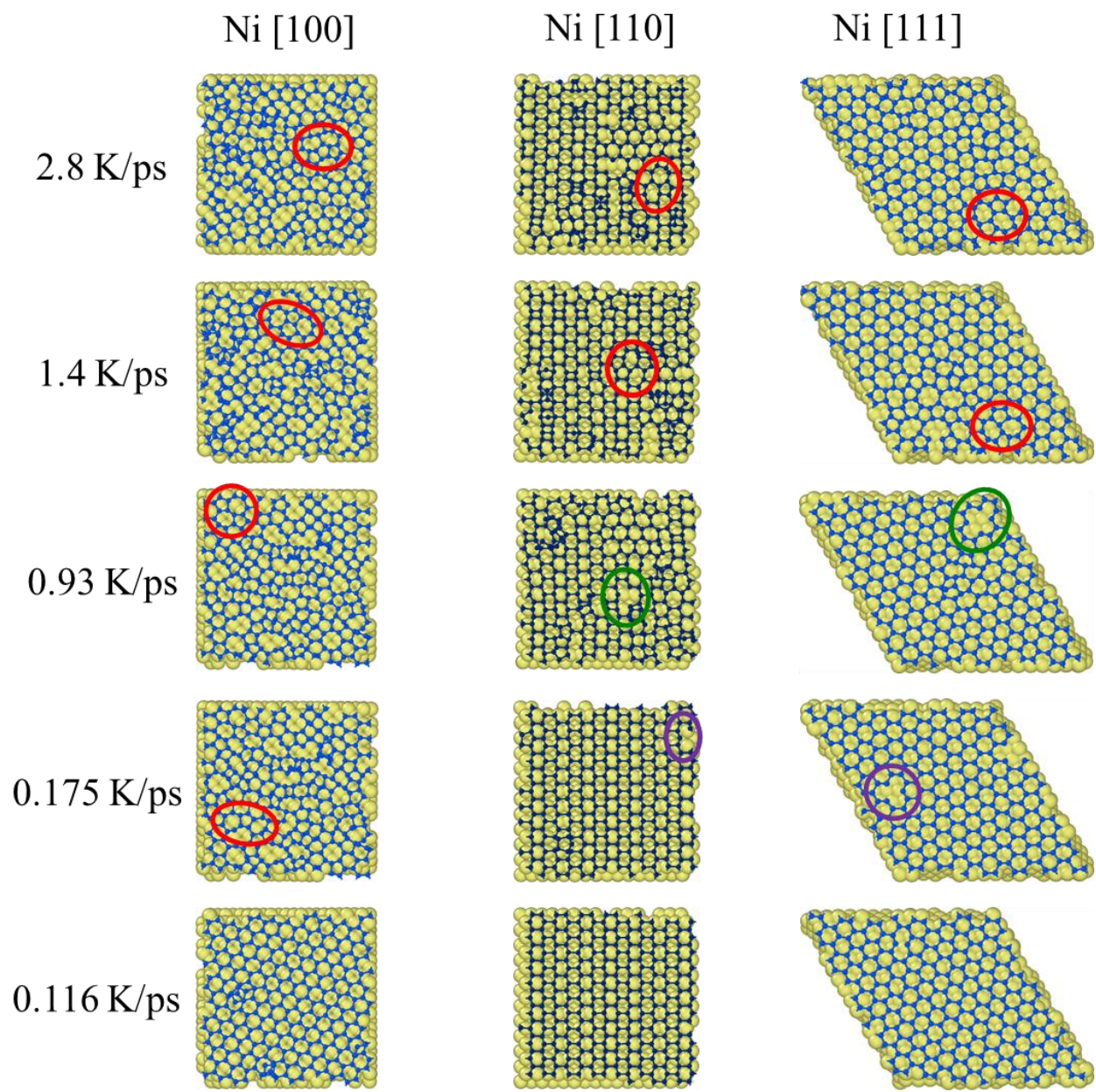

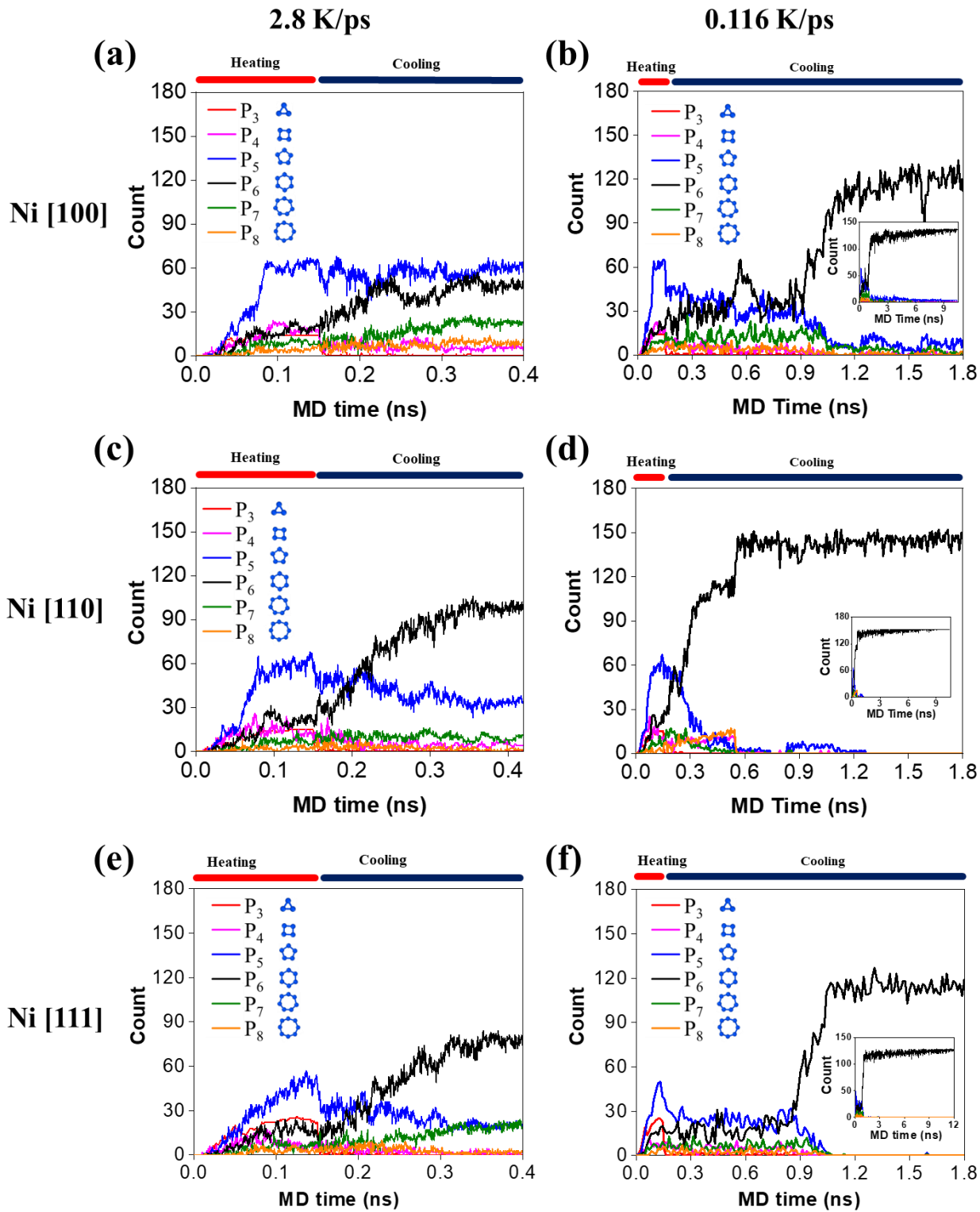

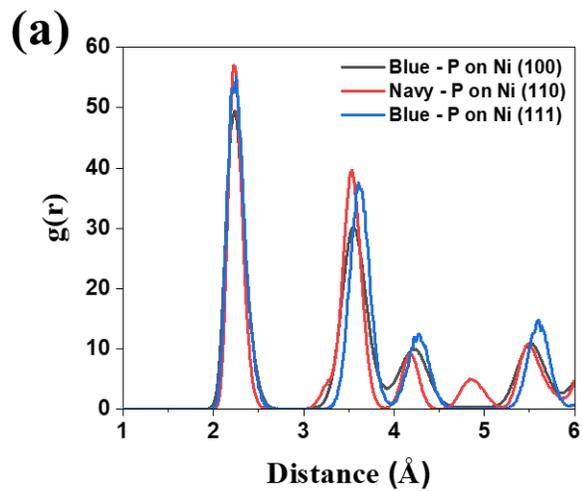 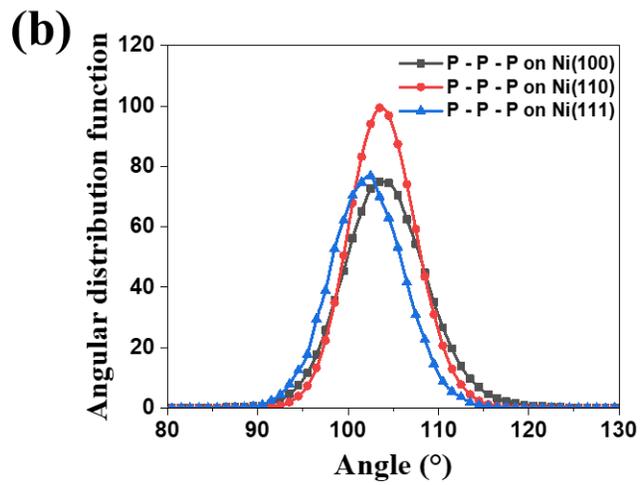

# SUPPLEMENTARY INFORMATION

## Towards large-area and defects-free growth of phosphorene on Nickel.

B. D. Tchoffo[1,+], I. Benabdallah[1*,+], A. Aberda[1], P. Neugebauer[2], A. Belhboub[3], A. El Fatimy[1 *]

[1] Institute of Applied Physics, Mohammed VI Polytechnic University, Lot 660, Hay Moulay Rachid, Ben Guerir, 43150, Morocco.

[2] Central European Institute of Technology, CEITEC BUT, Purkyňova 656/123, 61200 Brno, Czech Republic.

[3] École Centrale Casablanca, Ville verte, Bouskoura, Casablanca, Morocco.

[+] These authors contributed equally

*Corresponding author: ismail.benabdallah@um6p.ma, Abdelouahed.ELFATIMY@um6p.ma

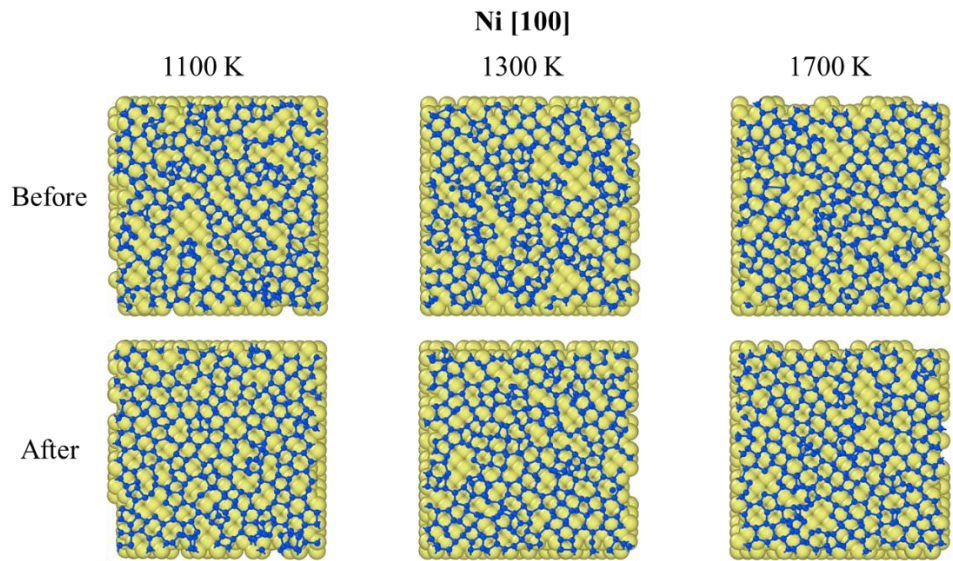
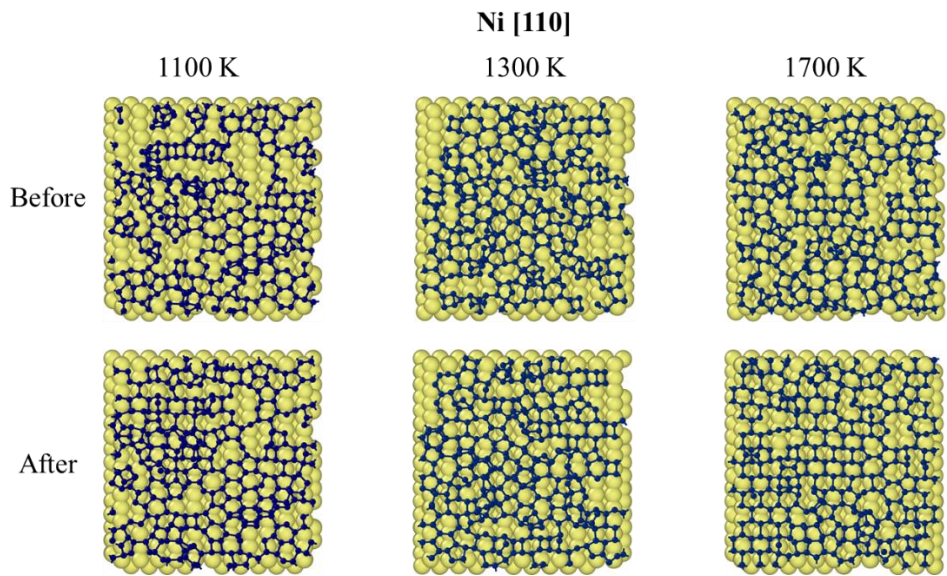

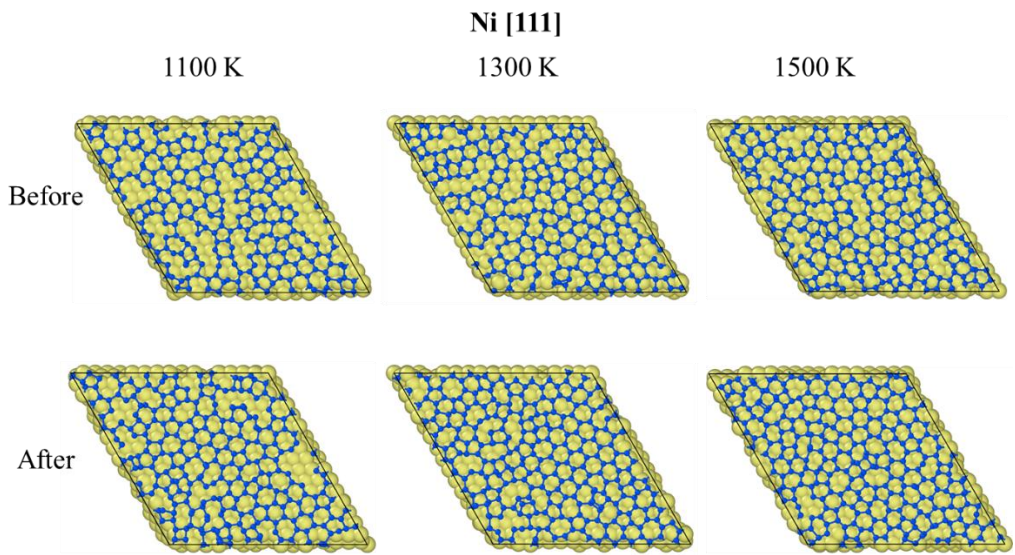

**Figure S1 :** (a) Top view of the final structure after heating to different temperatures before cooling

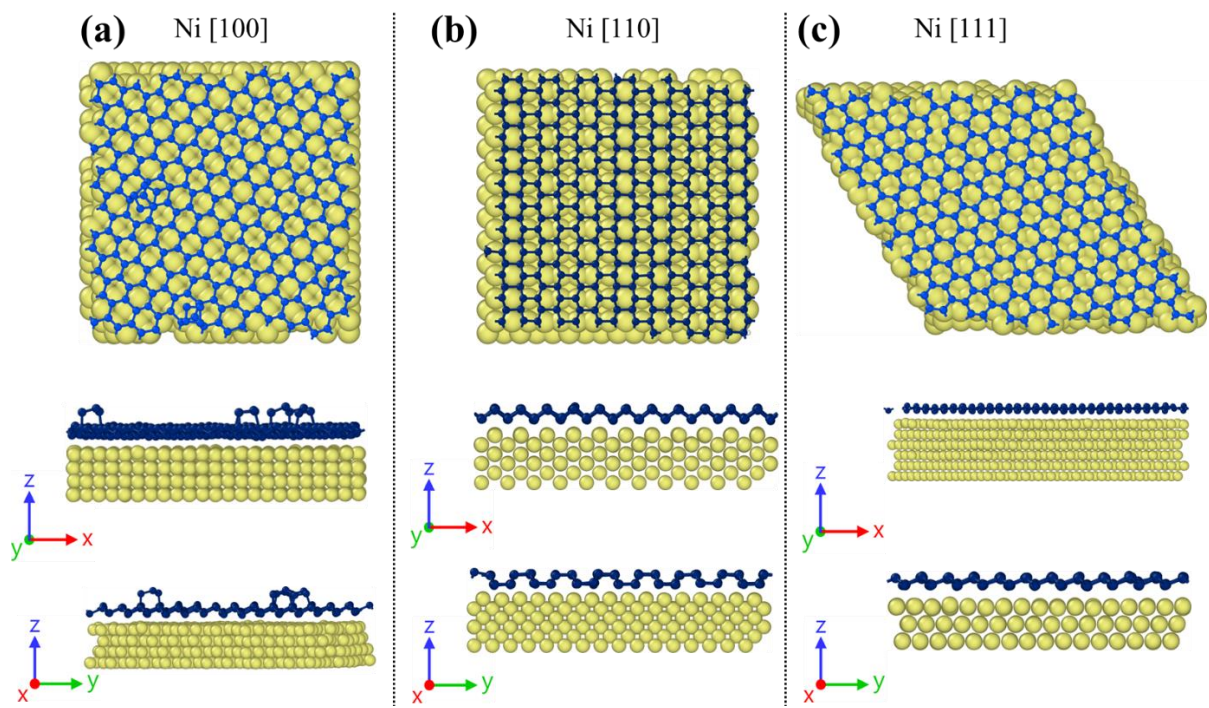

**Figure S2:** Top and Side view structures at 0.116 K/ps for Ni (100), (110) and (111).

Figures S2 (a), (b) and (d) shows the top and side views of the formed layer of β–P and γ–P on Ni (100), Ni (110) and Ni (111), respectively. The fact that the $P_5$ molecules play a crucial role as precursors for the formation of the phosphorene layer, the nature of the grown phosphorene also depends on the substrate surface. Therefore, this parameter allows a better selection and control for synthesizing to a precise allotrope.

**Supplementary Video 1** : Phosphorus atoms deposition on a Ni (111) at 1300 K without cooling.

**Supplementary Video 2** : Phosphorus atoms deposition on a Ni (111) with a fast cooling rate of 2.8 K/ps.

**Supplementary Video 3** : Phosphorus atoms deposition on a Ni (111) with a slow cooling rate of 0.116 K/ps.

**Supplementary Video 4** : Phosphorus atoms deposition on a Ni (110) with a slow cooling rate of 0.116 K/ps

## Python methods

Graph theory model has been used to count the number of n-membered rings formed by phosphorus atoms at each timestep. A graph of the system was constructed at each timestep. Phosphorus atoms were considered as nodes of the graph while their bonds represented the edges. The number of elementary cycles of n nodes in the graph corresponds to that of the n-membered rings formed by phosphorus atoms. We limited the count to rings formed only by 3 to 8 atoms. Note that the graph provides only a 2D representation of the system which can be inaccurate at the beginning of the simulation where some 3D configurations appear such as P4 molecules (tetrahedral) which are counted as P3 by our program.

To implement this algorithm, we used Python3 with the Networkx library to model the graphs.